\documentclass[aps,showpacs]{revtex4-1}
\usepackage{amsmath, amsthm, amssymb}
\usepackage{epsfig}
\usepackage{color}

\begin{document}

\title{The escape transition of a compressed star polymer:\\ Self-consistent
field predictions tested by simulation}

\author{Jaros{\l}aw Paturej$^{1,2}$, Andrey Milchev$^{3,4}$,
Sergei A. Egorov$^5$ and Kurt Binder$^4$}
\affiliation{$^1$Department of Chemistry, University of North Carolina, Chapel Hill, North
Carolina 27599-3290, USA
\\$^2$Institute of Physics, University of Szczecin
Wielkopolska 15, 70451 Szczecin, Poland
\\$^3$Institute of Physical Chemistry, Bulgarian Academy of Sciences, Sofia 1113,
Bulgaria
\\$^4$Institut f{\"u}r Physik, Johannes Gutenberg
Universit{\"a}t Mainz, Staudinger Weg 7, D-55099 Mainz, Germany
\\$^5$Department of Chemistry, University of
Virginia, Charlottesville, Virginia 22901, USA
}
\email{Corresponding author: paturej@live.unc.edu}
\begin{abstract}
The escape transition of a polymer ``mushroom'' (a flexible chain grafted to a
flat non-adsorbing substrate surface in a good solvent) occurs when the polymer
is compressed by a cylindrical piston of radius $R$, that by far exceeds the
chain gyration radius. At this transition, the chain conformation abruptly
changes from a two-dimensional self-avoiding walk of blobs (of diameter $H$, the
height of the piston above the substrate) to a ``flower conformation'', i.e.
stretched almost one-dimensional string of blobs (with end-to-end distance
$\approx R$) and an ``escaped'' part of the chain, the ``crown'', outside the
piston. The extension of this problem to the case of star polymers with $f$ arms
is considered, assuming that the center of the star is grafted to the substrate.
The question is considered whether under compression the arms escape all
together, or whether there occurs an arm by arm escape under increasing
compression. Both self-consistent field calculations and Molecular Dynamics
simulations are found to favor the latter scenario.
\end{abstract}


\maketitle

\section{Introduction}
\label{sec_intro}

\begin{figure}[htb]
\includegraphics[scale=0.35]{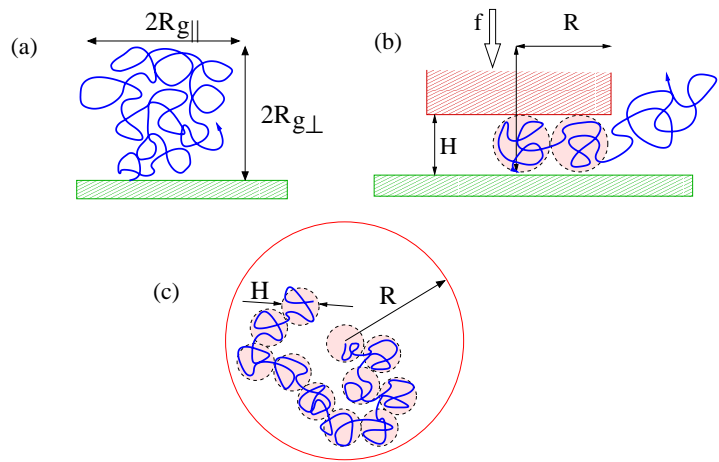}

\caption{\label{fig1} (a) Schematic plot of a ``mushroom'' (i.e., a chain
grafted with one end at a flat repulsive wall).  Linear dimensions of the coil
parallel $(R_{g\|}$) and perpendicular $(R_{g \bot}$) to the substrate are
indicated. Upon compressing the mushroom by a cylindrical piston of radius $R$
axially centered at the grafting site, the chain forms ``blobs'' of diameter $H$,
the height of the piston above the surface. Holding the piston at a chosen
height $H$ requires a force $\mathcal{F}$. The conformation of the chain may be
``escaped'' (part of the chain is not underneath the piston, case (b)) or
``imprisoned'' (the chain being fully underneath the piston, case
(c)). While cases (a) and (b) refer to a side view, case (c)
represents a top view.}
\end{figure}

During the last decades novel experimental techniques have been developed
allowing the observation and manipulation of single chains grafted to (or
adsorbed on) substrates \cite{1,2,3,4,5,6,7,8,9,10,11,12,13,14,15,16}.
A particular intriguing aspect is
the response of such grafted polymers to stretching or compression forces (for
excellent reviews of experimental work see e.g. \cite{17} and for the
theoretical aspects see \cite{18}). A particularly interesting phenomenon is
that such single polymer manipulations may induce unconventional conformational
transitions (see \cite{18} and refs. therein). In the present work we shall
focus on the so-called ``escape transition''
\cite{19,20,21,22,23,24,25,26,27,28,29,30,31,32}. Fig.~\ref{fig1} shows a
schematic sketch of the setup that is considered for the case of the escape of a
single flexible macromolecule grafted to a planar non-adsorbing surface, as
traditionally considered \cite{25}. Such a polymer, under good solvent
conditions, takes the so-called ``mushroom configuration'' \cite{33}, i.e. both

components of the gyration radius parallel $(R_{g\|})$ and perpendicular
$(R_{g\bot})$ of the chain scale with the number $N$ of effective monomeric units
as (cf. Fig.~\ref{fig1})
\begin{equation}
R_{g\|}\propto R_{g \bot} \propto a N^\nu
\label{eq1}
\end{equation}
where $a$ is the linear dimension of an effective monomeric unit, and $\nu$ is the
"Flory exponent" $\nu= 3/5$ (or, more precisely, $\nu \approx 0.588$
\cite{33,34}). When the mushroom is compressed by a cylindrical piston of radius
$R$ (we consider only the idealized case that the axis of the piston is
perpendicular to the surface, coincident with the $z$-axis through the grafting
site) to a height $H$ with $H \ll R_{g \bot}$ while $R \gg R_{g \|}$, one rather
obtains a ``pancake'' conformation, i.e. a two-dimensional self-avoiding walk of
``blobs'' of diameter $H$ \cite{33}. For a number $n$ of blobs, one has then

\begin{equation}
R_{g \|}= Hn^{3/4}
\label{eq2}
\end{equation}
since in $d=2$ dimensions the Flory exponent is $\nu_2= 3/4$ \cite{33}. The
standard scaling argument \cite{35} implies that inside a blob one still has the
same statistics as in the bulk ($d=3$ dimensions), so $H = a g^\nu$ if there are
$g$ effective monomeric units per blob. Since thus $g =(H/a)^{1/\nu} \approx
(H/a)^{5/3}$ and $n=N/g =N(H/a)^{-5/3}$, ignoring here and henceforth all
pre-factors of order unity in such scaling considerations, one obtains

\begin{equation}
R_{g \|} =N^{3/4} a(H/a)^{-1/4}\quad .
\label{eq3}
\end{equation}

The free energy cost for creating this confinement is simply the thermal energy
$k_BT$ times the number of blobs,
\begin{equation}
\Delta F (H)/k_BT = n=N(H/a)^{-5/3}\quad .
\label{eq4}
\end{equation}

When the chain is escaped one predicts instead a free energy cost independent of
$N$ \cite{19,20}
\begin{equation}
\Delta F (H)/k_BT = R/H\quad .
\label{eq5}
\end{equation}

Equating these two expressions yields a transition height

\begin{equation}
H_t/a= (Na/R)^{3/2}
\label{eq5}
\end{equation}
and this transition is accompanied by a jump in the force
$\mathcal{F} = -(\partial F/\partial H)_T$.

In the present paper, we shall be concerned with the escape transition when the
macromolecule is not a simple linear chain but has a star polymer architecture
\cite{36,Egorov,CNL,CNL1,CNL2}.
In the limit where $H$ is much smaller than the radius of a free
star, the configuration of a star polymer with $f$ arms confined into a slit of
width $H$ is essentially a quasi-two-dimensional star polymer, where each arm
occupies a slice with an angle $2 \pi/f$ cut from a cylinder of height $H$ and
radius $R_{\|}(f)$ with \cite{37,38,39,40}

\begin{equation}
R_\| (f) = f^{1/4}(H/a)^{-1/4} N^{3/4}.
\label{eq7}
\end{equation}

The free energy in this case simply is \cite{40}

\begin{equation}
\Delta F (H,f)/k_BT= fN(H/a)^{-5/3}\quad .
\label{eq8}
\end{equation}

Comparing Eqs.~(\ref{eq4})-(\ref{eq8}), one simply notes that the free energy is
additive in its contributions from the individual arms, which also have the same
number $n$ of blobs of diameter $H$ as for a linear chain, in this limit of
strong confinement.

When the star polymer is not compressed by a plate of infinite lateral extent
but by a cylindrical piston of finite radius $R$, as in Fig.~\ref{fig1}, an
escape transition for a star polymer also becomes conceivable, and the question
that immediately comes to mind is: will all arms escape at this transition, or
will there be a sequence of $f$ transitions, due to arm-by-arm escape? Sevick
\cite{41} suggested that the latter scenario applies, based on a simple Flory
theory treatment \cite{33}.

In the present work, we reconsider this problem, giving for the first time a
more detailed self-consistent field treatment (Sec.~\ref{sec_scft}) and Molecular Dynamics
Simulations (Sec.~\ref{sec_MD}). Note that for finite chain length $N$ the
escape transition is not a sharp first-order transition but rather rounded by
finite size effects \cite{25,26,32}, and in view of multiple transitions located
close by to each other it is conceivable that the predicted singular behavior is
completely washed out in cases of practical interest. Thus, it is important to
study this problem beyond the simple level of Flory theory.
Finally, Sec.~\ref{sec_summary} presents discussion and concluding
remarks. Note that Molecular Dynamics and Self-Consistent Field Methods are
complementary: Molecular Dynamics is more accurate, taking all statistical
fluctuations into account and, since it uses continuum models, is also more
realistic. The Self-Consistent Field method, however, which uses lattice
formulation, implies mean-field approximations and hence is less accurate
albeit it allows to treat much longer chains and stars with more arms, which is
 important for the present problem.

\section{Self-Consistent Field Theory}
\label{sec_scft}

As is well-known, the self-consistent field (SCF) approach \cite{42} takes excluded
volume interaction into account only via a mean field approximation, like Flory
theory \cite{33} does, but unlike the latter it considers explicitly the
non-uniformity of the monomer density distributions and deals with effects due
to finite chain length: as is necessary for the present problem. Using the
lattice discretization as developed by Fleer et al. \cite{42}, one can derive
SCF equations for the volume fraction profiles of all monomeric
species, putting the effective segment size $a$ equal to the lattice spacing
(which is our unit of length in this section). The SCF
equations are non-linear and need to be solved numerically by an iteration
procedure. For this purpose, it is advantageous to take the symmetry of the
problem into account (in the present case, one can invoke a cylindrical symmetry
around the axis of the cylindrical piston, and formulate the problem in
cylindrical coordinates, as described in \cite{43} in another context). Grafting
is achieved by pinning an end-segment of each arm of the star to the core.
Technical details of the method were described in the literature in detail
\cite{40,43} and hence will not be repeated here.

\begin{figure}[htb]
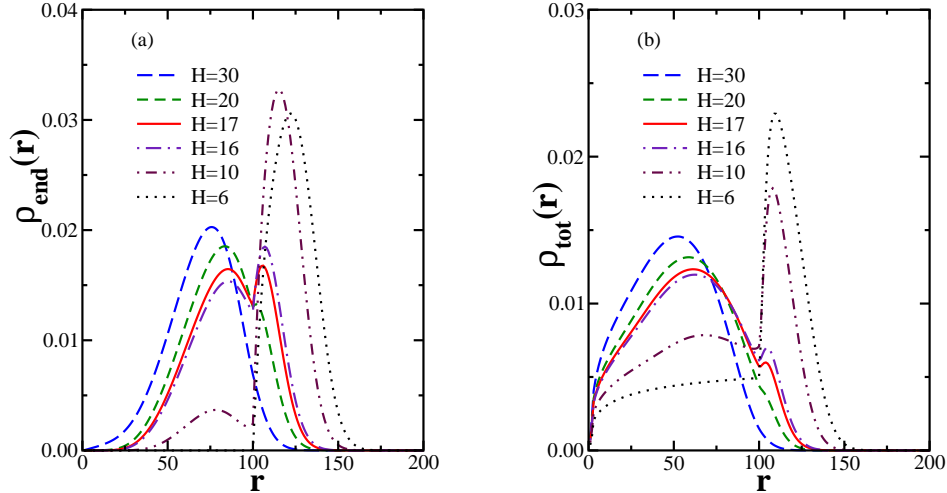

\vspace{0.7cm}
\includegraphics[scale=0.3]{fig2a.eps}
\hspace{1.0cm}
\includegraphics[scale=0.3]{fig2b.eps}
\vspace{0.7cm}

\caption{\label{fig2} (a) SCF results for the radial density
profiles $\rho_{end}(r)$ of the end segments for a star polymer grafted with its
core at $r=0$, $z=0$, for the case $f=10, N=1000, R=100$, and six different
piston heights $H=6,10,16,17,20$ and 30, as indicated. All lengths are measured in
units of the lattice spacing $a$. (b) same as (a) but  for the radial density
profiles $\rho_{tot}(r)$ of all segments. Note that both distributions
are normalized to unity according to
$\int\rho_{end}(r) dr = \int\rho_{tot}(r) dr= 1$ and thus incorporate
a weighting factor $2\pi r$ already in their definition.}
\end{figure}

We consider a star polymer with $f=10$ arms of length $N=1000$ tethered by its
core segment to a flat surface. In a first step, we compressed this star polymer
by a cylindrical piston of radius $R=100$. In Fig.~2a we show results
for the radial density profiles of the free chain ends of the arms (not
distinguishing yet to which arm the ends belong). One sees that for $H=30$ and
$H=20$ the distribution has a single peak at $r=r_{max}^{(1)} < R$; as $H$
decreases $r_{max}$ increases, as expected. For $H=16$ and $H=17$ a double-peak
structure with two peaks at $r_{max}^{(1)} <R, \quad r_{max}^{(2)} >
R$ of comparable height is seen, while for $H=10$ and $H=6$ the peak at $r_{max}^{(2)}$ clearly
dominates. Thus, these results give clear qualitative evidence that for the
chosen parameters an escape transition of star polymers can be observed.
In Fig.~2b we show results
for the radial density profiles of all the monomers of the star for the same six
values of the piston height as in Fig.~2a.

\begin{figure}[htb]
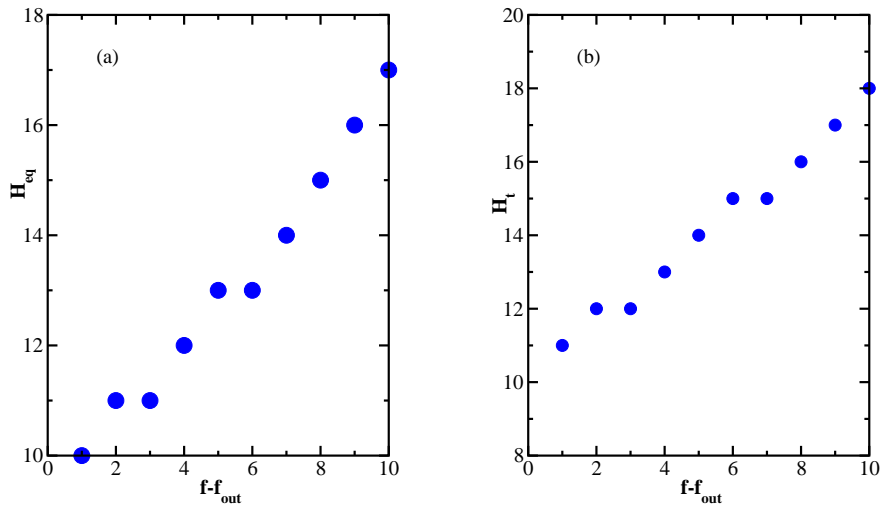

\vspace{0.7cm}
\includegraphics[scale=0.3]{fig3a.eps}
\hspace{1.0cm}
\includegraphics[scale=0.3]{fig3b.eps}
\vspace{0.7cm}

\caption{\label{fig3} (a) SCF results for $H_{eq}(f_{out})$, the
piston height at which the two peaks of $\rho_{end}(r)$ in Fig.~2 have equal
heights, versus $(f-f_{out})$, where $f$ is the total number of arms and
$f_{out}$ is the number of arms that are constrained to have their
end segments out of the piston.
(b) Height $H_t$ at which the escape transition of the first
arm (from a fully ``imprisoned'' state of confinement) occurs
plotted vs. $f-f_{out}$;
dots are from the locations of the ``jumps'' in Fig.~\ref{fig5}c.}
\end{figure}

In order to turn to the issue of the arm-by-arm escape predicted by the Flory
theory \cite{41}, we have adopted the following strategy. We sequentially
constrain $f_{out}=1,2,3,\ldots$ star arms to the end segment outside of the
piston, and determine the corresponding piston height $H_{eq}(f_{out})$ where
the two peaks in the end segment distribution have equal heights. In
the left panel of Fig.~\ref{fig3} we plot $H_{eq}(f_{out})$ versus $f-f_{out}$. One sees that
$H_{eq}(f_{out})$ shows a monotonic decrease with increasing $f_{out}$,
which may be taken as an indirect evidence for arm-by-arm escape.
\begin{figure}[htb]
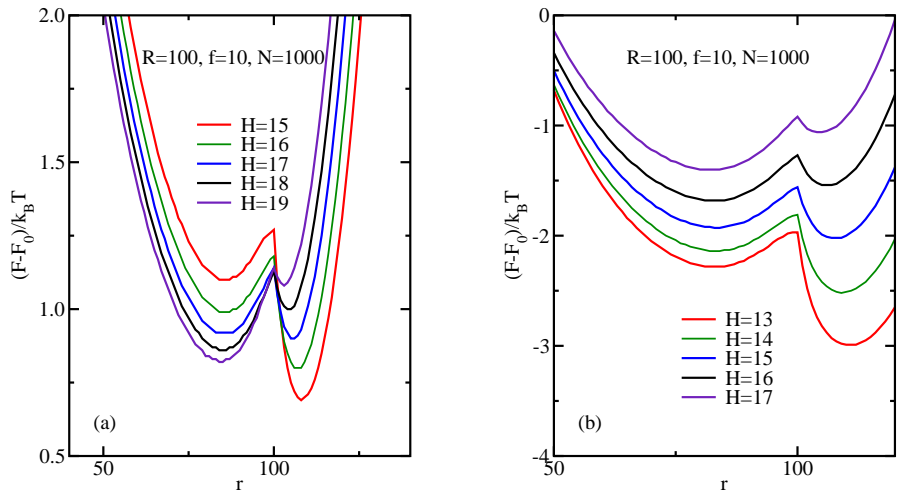

\vspace{0.7cm}
\includegraphics[scale=0.3]{fig4a.eps}
\hspace{1.0cm}
\includegraphics[scale=0.3]{fig4b.eps}
\vspace{0.7cm}

\caption{\label{fig4} (a) Free energy function $F(r)$ relative to its reference
value $F_0$ for the star polymer plotted as a function of the position $r$ of
the end monomer of one arm having its end monomer constrained to be at a radial
distance $r$ from the axis. Five different heights $H=15,16,17,18$ and 19 are
included, as indicated. All data refer to the choice $R=100, f=10,$ and N =
1000. (b) Same as (a), but with 3 arms escaped. Only choices of $H$ near the
transition value $H$ are shown, namely $H=13,14,15,16$ and 17.}
\end{figure}

While it is known that the absolute values for the free energy of polymers
predicted by the SCF theory often are unreliable \cite{40}, we
nevertheless expect that constructing a Landau free energy function similar to
the spirit of the study of the escape transition of single chains \cite{32} will
give a useful first orientation (Fig.~\ref{fig4}a).
Indeed, we see that $F(H,r)$ as function of the position $r$ of one constrained
end-monomer develops a double well structure, with a barrier occurring at $r=R$;
the scale of this barrier is only of order unity, however. Similar low barriers
were also found for the escape transition of linear chains \cite{32}. Thus,
Figs.~\ref{fig3} and~\ref{fig4}a imply that although we expect that indeed
arm-by-arm escape will occur in the thermodynamic limit (note that $N
\rightarrow \infty$ and $R\rightarrow \infty$ must be taken together, keeping
the ratio $Na/R$ fixed, cf.~Eq.~(\ref{eq6}), to obtain a sharp phase transition
characterized by a truly singular behavior), for star polymers with physically
realistic choices of parameters the transitions must be strongly rounded, and
states with $f_{out}=1,2,3,\ldots$ escaped arms have strongly overlapping
distributions of all the observable properties. Along these lines,
Fig.~\ref{fig4}b compares the free energy of the imprisoned and escaped stars
for different choices of $H$ in the vicinity of the transition at $H_t$, where
one arm escapes, while two other arms are constrained to have their end-segment
anywhere outside of the piston.

\begin{figure}[htb]
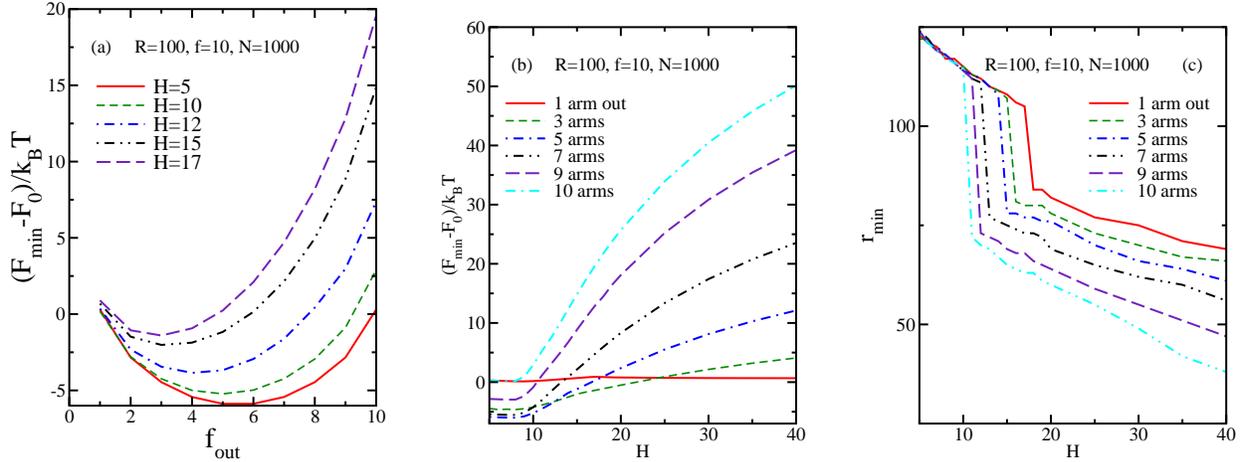

\vspace{0.7cm}
\includegraphics[scale=0.27]{fig5a.eps}
\hspace{0.5cm}
\includegraphics[scale=0.27]{fig5b.eps}
\hspace{0.5cm}
\includegraphics[scale=0.27]{fig5c.eps}
\vspace{0.7cm}

\caption{\label{fig5} Free energy in its minimum $F_{min}$ (cf.~Fig.~\ref{fig4})
relative to the reference value $F_0$ for stars with $f=10$ arms of length
$N=1000$, plotted versus the number $f_{out}$ of arms constrained to be escaped,
for different choices of $H$ (a), and alternatively plotted as function of $H$,
for different choices of $f_{out}$ (b). The lowest branches (in between the
crossing points) clarify the range for which every value of $f_{out}$ yields the
stable state, while branches with higher free energy are metastable. Part (c)
shows the variation of the minimum position $r_{min}$ of $F_{min}$, for several
choices of $f_{out}$, as indicated.}
\end{figure}

Next we consider the variation of the free energy with the number of
arms constrained to be outside the piston, $f_{out}$  (Fig.~\ref{fig5}).
Plotting it versus $f_{out}$ for different values of $H$, Fig.~\ref{fig5}a, we
recognize that e.g., for $H=17$ and $H=15$ this minimum occurs at $f_{out}=3$,
while for $H=12$ it occurs for $f_{out}=4$, for $H=10$ for $f_{out}=5$ and for
$H=5$ for $f_{out}=6$. In Fig.~\ref{fig5}b, the results are plotted
alternatively as function of $H$, for different choices of $f_{out}$. The lowest
branches (in between the crossing points) clarify the range for which every
value of $f_{out}$ yields the stable state, while branches with higher free
energy are metastable. Finally, Fig.~\ref{fig5}c shows the variation of the
minimum position $r_{min}$ of $F_{min}$, for several choices of $f_{out}$, as
indicated.
From the locations of the ``jumps'' in Fig.~5c,
one can determine the height $H_t$ at
which the escape transition of the first arm (from a fully `imprisoned'' state
of confinement) occurs. This height is plotted vs. $f-f_{out}$  as
dots in the right panel of Fig.~\ref{fig3}, which shows behaviour
qualitatively similar to the one seen in the left panel of the same Figure.

\begin{figure}[htb]
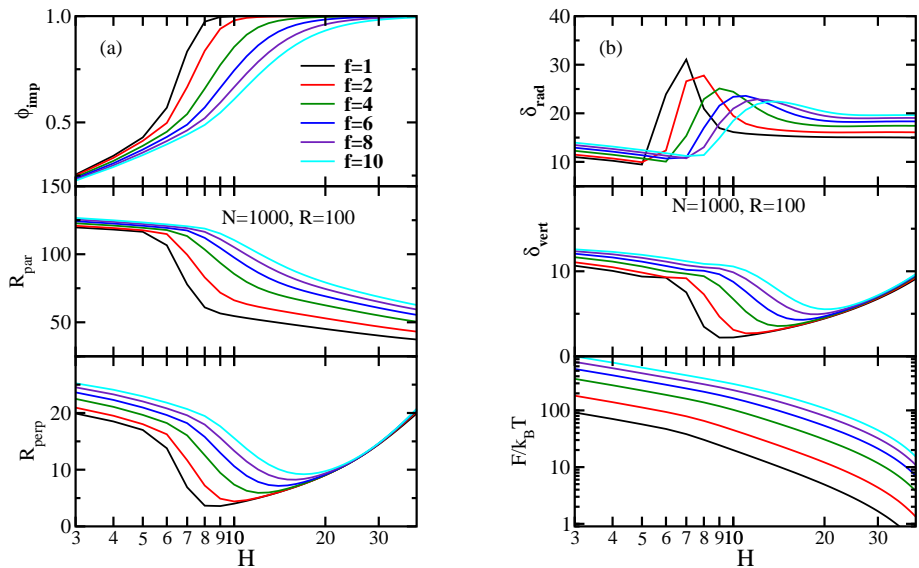

\includegraphics[scale=0.3]{fig6a.eps}
\hspace{1.0cm}
\includegraphics[scale=0.3]{fig6b.eps}
\vspace{1.0cm}

\caption{\label{fig6} (a) Fraction of imprisoned monomers, $\phi_{imp}$, (upper panel), and
parallel, $R_{g \|}$, (middle panel) as well as perpendicular,
$R_{g\bot}$,
(lower panel) gyration radius component of the grafted star polymer, plotted
vs. the piston height $H$. Various numbers of arms $f$ are included, as
indicated; all data are for the case $R=100,\; N = 1000$. (b) Fluctuations of
the free end-monomer in the radial, $\delta_{rad}$, (upper panel) and
vertical, $\delta_{vert}$, (middle panel)
directions. Lower panel: dimensionless compression free energy, $F/k_BT$, as a function
of the piston height. The total free energy has a smooth, nonsingular variation,
due to the rounding of the transition.}
\end{figure}

In Fig.~\ref{fig6} we present SCF results for several structural and
thermodynamic properties of compressed stars as a function of the piston height
$H$, while the number of arms $f$ is varied from $1$ (linear  chain) up to $10$;
in all cases the arm length is $N=1000$, and the piston radius is $R=100$. In
particular, the upper panel of Fig.~\ref{fig6}a shows the variation with the
piston height $H$ of the ``order parameter'' (the fraction of imprisoned
monomers, $\phi_{imp}$), while middle and lower panels  present the parallel
and perpendicular gyration radius component of the star, respectively. From the
upper panel, one sees that the transition is sharpest for the single chain
$(f=1)$, and becomes more and more blurred as the number of arms increases. Note
the non-monotonic increase of $R_{g \bot}$ for small $H$: this reflects the
contribution from the blob outside of the piston (Fig.~\ref{fig1}) which
increases in size as $H$ decreases.

The upper and middle panels of Fig.~\ref{fig6}b present the free end-monomer
fluctuations in the radial and vertical directions, respectively. One sees that the
fluctuations in the radial direction display a pronounced maximum as a
function of the piston height $H$, whose height and sharpness both
decrease with increasing number of arms, while its location moves to
larger values of $H$, as one would expect. By contrast, fluctuations
in the vertical direction show a minimum as a function of $H$, which
is presumably related to the minimum observed in the vertical
(perpendicular) component of the gyration radius of the grafted star polymer.
The lower panel of Fig.~\ref{fig6}b shows compression free energy as a
function of the piston height; it has a smooth, nonsingular variation, due to
the rounding of the transition.

Referring back to Fig.~\ref{fig4}, we also note that the barrier between the two
minima, separating the state with all arms being still confined (the left
minimum) and the first arm being already escaped (the right minimum)
decreases with increasing number of arms, and at the same time, the height $H_t$
at which this transition occurs increases (right panel of
Fig.~\ref{fig3}). This behavior is
qualitatively easy to understand: the radius $R_{\|}(f)$ of a fully confined
star increases with $f$ (Eq.~(\ref{eq7})). On the other hand, it is
not straightforward to predict this behavior from the theory: one could argue
that the free energy of a compressed star with $f$ arms is simply additive with
respect to the contributions of the $f$ arms. If we would assume that this is
still true when one of the arms is escaped, we would predict that the free
energy then is

\begin{equation}
\Delta F_{1 \; arm \; esc} (H,f)/k_BT = (f-1)N(H/a)^{-5/3} +R/H,
\label{eq9}
\end{equation}

where we have taken Eq.~(\ref{eq5}) again to describe the contribution of the
escaped arm to the free energy, and the remaining $f-1$ arms that are still
imprisoned yield the same contribution as an imprisoned star with only $f-1$
arms. However, when Eq.~(\ref{eq9}) would hold, the transition between the state
with $f$ arms confined (Eq.~(\ref{eq8})) and with $(f-1)$ arms confined
(Eq.~(\ref{eq9})) would still take place at $H_t/a$ as given by Eq.~(\ref{eq6}),
i.e. a result independent of $f$. However, such a result would be at variance
with Fig.~\ref{fig3}.
Remember that the escape transitions of the successive arms that leave the region underneath
the piston are all strongly rounded, and thus different criteria to
locate the transitions
give slightly different results,
as expected;
but the range of $H$ over which the variation of the location of the
transition with $f$ changes is much larger
than the extent of the rounding.

\section{Molecular Dynamics Simulations}
\label{sec_MD}

We study a coarse-grained model \cite{44} of star polymers with $f$ arms
containing $N=150$ effective monomers each. This model has been studied
extensively before, in bulk solution under good solvent conditions
\cite{44}, and for stars strongly adsorbed on a surface \cite{Egorov}
or under confinement in planar slits \cite{40}; hence we summarize here only a few
details of this model. The effective monomers interact with the repulsive part
of the (shifted and truncated) Lennard-Jones potentials
\begin{equation}
V(r) = 4 \epsilon [(\sigma/r)^{12}-(\sigma/r)^6 +1/4]\; \quad r <
r_c = 2^{1/6} \sigma \; ;
\label{eq10}
\end{equation}
$V(r>r_c) \equiv 0$, and the strength $\epsilon \equiv 1$ and range
$\sigma \equiv 1$ of this potential are taken
as units of temperature and length, respectively. Bonded monomers
experience in addition the ``FENE potential'' \cite{40,41},
\begin{equation}
V_{FENE} (r) = - 0.5 k r_0^2 \ln [1-(r/r_0)^2], \quad k = 30 \epsilon/\sigma ^2,
\; r_0 = 1.5 \sigma
\label{eq11}
\end{equation}

As usual, Molecular Dynamics \cite{44,45} simulations are performed, using the
Velocity Verlet algorithm to integrate the (Newtonian) equations of motion, to
which a friction term plus a random force (related to the friction coefficient
by the fluctuation-dissipation relation) is added. Using $m=1$ as mass for the
particles, time is measured in units of $\tau _{MD} = (m \sigma
^2/\epsilon)^{1/2}$, and the integration time step then was taken 0.002
$\tau_{MD}$ (using a friction coefficient $\gamma = 0.25$). Runs with up to
10$^7$ MD steps were performed, using typically $f=6$ and $N=150$ while the
piston radius $R$ and height $H$ above the substrate needed to be varied.
\begin{figure}
\includegraphics[scale=0.45]{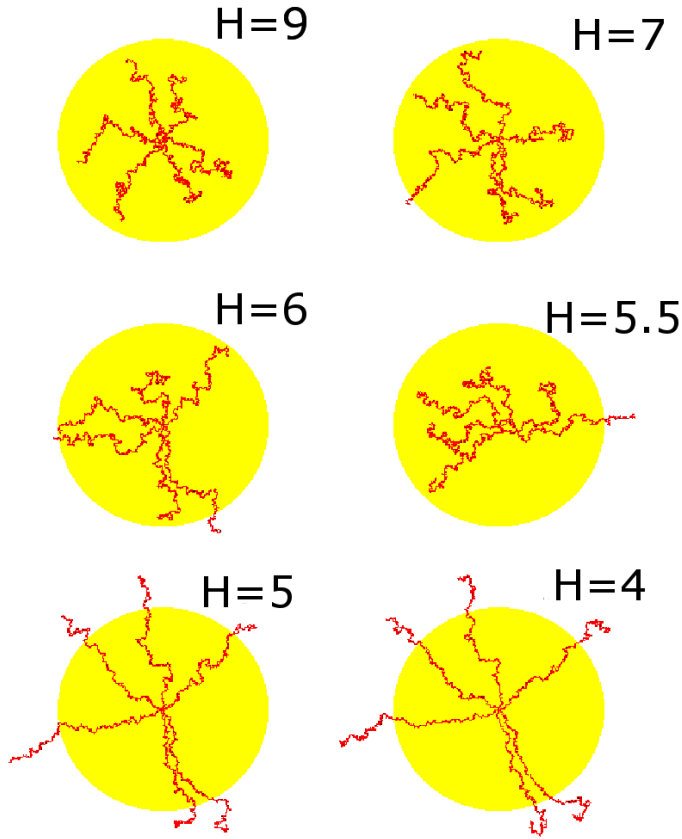}
\vspace{0.7cm}

\caption{\label{fig7} Snapshot pictures of a $6$-arm star (top view) of a star
polymer with N=150 grafted underneath a piston of radius $R=50$ for $6$ choices
of $H$, as indicated.}
\end{figure}

Fig.~\ref{fig7} shows typical snapshot pictures for $R=50$ and several choices
of the height $H$ of the piston above the substrate. For $H=9$ the star
typically is still completely imprisoned; for $H=7$ down to $H=5.5$, one
occasionally observes that one of the arms ``tries to escape'', but typically
there occur a lot of fluctuations, one arm that has escaped retracts again, and
another arm escapes. For still smaller $H$, however, such as $H=5$ or $4$, all
arms have escaped already.
\begin{figure}[htb]
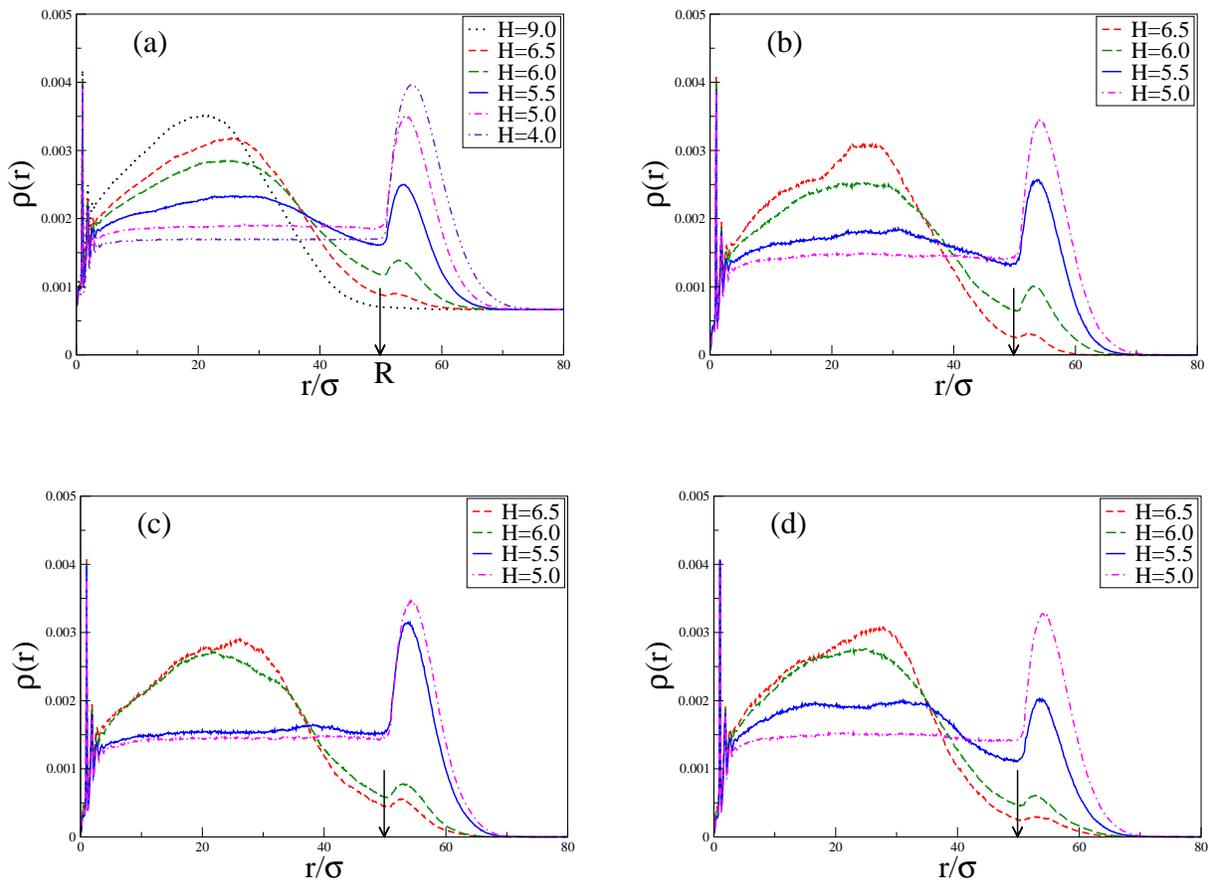

\includegraphics[scale=0.3]{fig8a.eps}
\hspace{0.7cm}
\includegraphics[scale=0.3]{fig8b.eps}
\vspace{1.2cm}

\includegraphics[scale=0.3]{fig8c.eps}
\hspace{0.7cm}
\includegraphics[scale=0.3]{fig8d.eps}
\caption{\label{fig8} (a) Radial density probability distribution $\rho(r)$
versus dimensionless distance (measured from the grafting point) $r/\sigma$,
averaged over all $6$ arms of a star polymer with $f=6$ and $N = 150$. Several
piston heights $H/\sigma$ are included as indicated. (b) - (d) Same as (a), but
for the individual arms $1,\; 3,$ and $5$. The rim of the piston with radius
$R=50$ is indicated by an arrow.}
\end{figure}

Fig.~\ref{fig8}a shows the resulting radial monomer density distribution
$\rho(r)$,
taking all monomers into account. One sees that for small $r$ ($r/\sigma \leq 4$)
characteristic oscillations occur, describing a radial layering of the effective
monomers around the star center. Of course, this is a model-specific detail,
which is of little interest in a scaling context.
Apart from this special feature, there is a striking qualitative
similarity with the corresponding results of SCF theory (Fig.~2b).
For $H/\sigma = 9$ then
$\rho(r)$ exhibits a peak near $r/\sigma \approx 20.$ So one would consider this
state, in the standard terminology of escape transitions, as an ``imprisoned''
star, but one should note that a small fraction of monomers is already in
the region outside of the piston. For $H/\sigma = 6$ we see already a double
peak distribution: the maximum of the escaped part being at $r/\sigma \approx
52$, and the maximum of the ``imprisoned'' part at $r/\sigma \approx 25$. This
maximum clearly is still the dominating one, while for $H=5$ the situation has
reversed, the maximum representing the escaped part dominates. At very large
compression $(H/\sigma = 4)$ the linear variation of $\rho(r)$ with $r$
indicates that the imprisoned part consists just of radially stretched strings
of blobs.

The bimodal character of the distribution $\rho(r)$ for intermediate values of
$H$ indicates that this escape transition can be viewed as a finite size rounded
first order transition; but clearly the data are insufficient to distinguish a
single transition from a series of transitions located close by each other.
Figs.~8b,c,d show three typical examples for distributions recorded for single
arms (only for $H/\sigma =5, 5.5, 6$ and 6.5, respectively). One sees huge
fluctuations from arm to arm: some are almost fully escaped in the transition
region, some still ``imprisoned''. Of course, if the running time of the
simulation were orders of magnitude larger, all arms would yield an identical
distribution. Thus, while the simulations are certainly suggestive of a sequence
of transitions, starting with a transition where a single arm escapes, a
substantially larger computational effort (hardly feasible at present) would be
needed to prove such an arm-by-arm escape.

\begin{figure}
\includegraphics[scale=0.35]{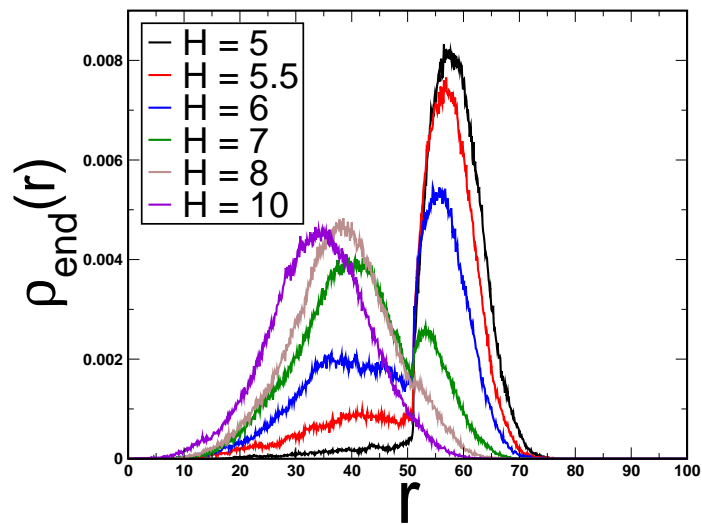}
\vspace{0.7cm}
\caption{\label{fig9} Radial density probability distribution of end-monomers
$\rho_{end}(r)$
versus dimensionless distance (measured from the grafting point) $r/\sigma$,
averaged over all arms of a star polymer with $f=6$ and $N = 150$. Several
piston heights $H/\sigma$ are included as indicated; the piston radius is $R=50$.}
\end{figure}

In Fig.~\ref{fig9} we show simulation results
for the radial density profiles $\rho_{end}(r)$ of the  end-monomers of the star arms (not
distinguishing to which arm the ends belong). One sees that for $H=10$ and
$H=8$ the distribution has a single peak at $r=r_{max}^{(1)} < R=50$; as $H$
decreases $r_{max}$ increases, as expected. For $H=6$ and $H=7$, a double-peak
structure with two peaks at $r_{max}^{(1)} <R, \quad r_{max}^{(2)} > R$ of comparable
height is observed, while for $H=5$  and $H=5.5$ the peak at $r_{max}^{(2)}$ clearly
dominates.

\begin{figure}
\vspace{1cm}
\includegraphics[scale=0.25,angle=270]{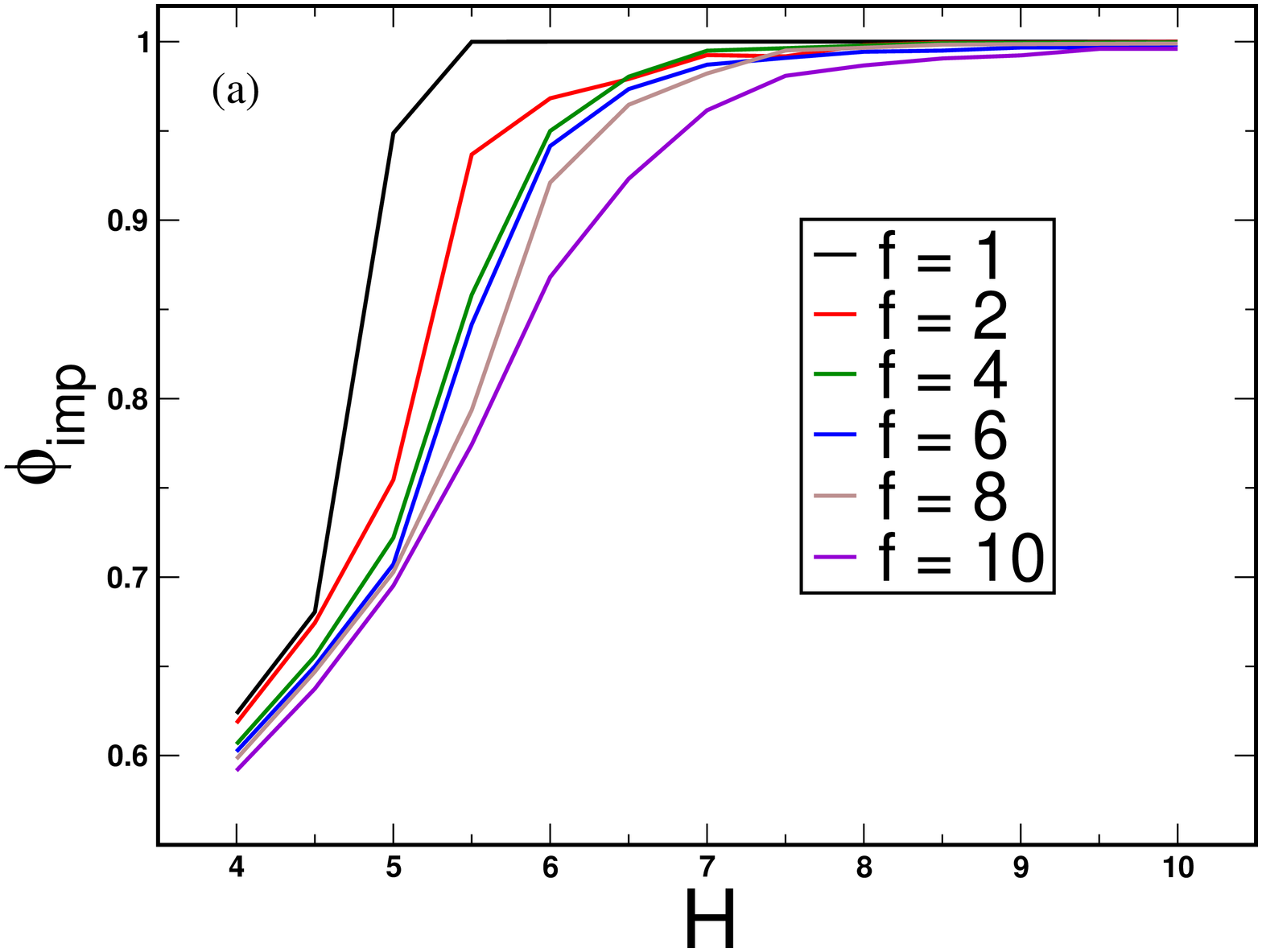}
\vspace{1cm}

\includegraphics[scale=0.25]{fig10b.eps}
\vspace{1cm}

\includegraphics[scale=0.25]{fig10c.eps}

\caption{\label{fig10} (a) Fraction of imprisoned monomers of the grafted star polymer, plotted
vs the piston height $H$; various numbers of arms $f$ are included, as
indicated; all data are for the case $R=50,\; N = 150$.
(b) Parallel component of the squared gyration radius vs the piston height $H$.
(c) Perpendicular component of the squared gyration radius vs the piston height $H$.}
\end{figure}

Finally, in Fig.~\ref{fig10} we display simulation results for the
same three observables as shown earlier in Fig.~6a for the SCF
calculations, i.e. fraction of imprisoned monomers (Fig.~10a), and
both parallel (Fig.~10b) and perpendicular (Fig.~10c) components
of the squared gyration radius of the grafted star polymer. All three
observables are plotted as a function of the piston height $H$ and
display the same qualitative behavior as seen earlier in the SCF
results shown in Fig.~6a.

\section{Discussion and Concluding Remarks}
\label{sec_summary}

In recent years, the effect of chemical architecture of macromolecules
(block and graft copolymers, stars, dendrimers, etc.) has become a
subject of great interest,
since these macromolecules can serve as building blocks of various
novel materials.
Parallel to this development, the mechanical manipulation of
macromolecular objects by external devices
(AFM tips, optical and magnetic tweezers) has been developed towards
maturity, and has yielded a lot
of insight into the function of biological molecules grafted to
substrates such as biomembranes.
In such contexts, it is an interesting variation to consider a grafted
star polymer and study its
response to mechanical compression, and the escape transition that
becomes possible when the
mechanical compression acts only over the area of a circle of radius $R$
(Figs.~\ref{fig1},~\ref{fig7}), so that some (or all) arms of the star
may avoid too strong compression forces by forming a ``flower (stem
plus crown)'' conformation
(i.e., a stretched string of blobs extends to the boundary of the
piston, so that many monomeric units are in a large ``blob'', the
``crown'', outside of the
confining piston). While we are not aware that such an experiment
already have been performed,
we feel that such an experiment should be feasible, and could yield
interesting insight
into the interplay of various entropic forces controlling the
conformation of confined macromolecules.
Note that instead of an ordinary star polymer one can also graft a
spherical polymer brush,
containing a (spherical) nanocolloid as a core, to which $f$ long
chains (with radii larger than
the radius of the nanocolloid) are grafted. We hope that our
simulation study will stimulate
such experiments.\\

Using both SCF theory and Molecular-Dynamics
simulations,
we have shown that for typical choices of parameters, the transitions
where one arm after the other escapes the confinement are not sharply
separated from each other,
but occur rather gradually
(Figs.~\ref{fig2},~\ref{fig3},~\ref{fig6}, and~\ref{fig10}).
While the general feature of escape transitions  for linear chains and for
grafted star polymers are qualitatively similar, the distinctive feature of
star polymers is that in the long star limit a sequence of arm-by-arm
transitions emerges.
\\

Only when one extracts the coarse-grained free energy functions
(Figs.~\ref{fig4},~\ref{fig5})
that correspond to configurations constrained such that $f_{\rm out}$
arms are escaped
and $f-f_{\rm out}$ arms are not yet escaped, one can verify that in
the thermodynamic
limit a sequence of first-order transitions due to arm-by-arm escape
will result;
for typical choices of parameters,
the barriers between the states do not exceed the thermal energy.
An interesting extension concerns the competition between adsorption of star
polymers and escape when the confining surfaces exhibit an attractive
interaction with the monomers. However, this problem must be left for a future
study.
\\

While Molecular Dynamics simulations, in principle, provide exact
statistical mechanics
for the chosen model system (apart from statistical errors!),
the computational effort for parameters of interest is still very
large,
precluding a systematic variation of all these parameters.
Such a variation is considerably easier for the SCF theory:
it is well established that the theory works well for dense polymer melts,
its accuracy for fairly dilute systems has been rather uncertain.
Thus, it is gratifying that for the present problem the simulations
reveal a striking
qualitative similarity with the SCF results. Of course, a
quantitative agreement cannot occur,
due to the differences between the lattice model of SCF and continuum
model used in the simulations.\\

\underline{Acknowledgement}: One of us (A.M.) received partial support
by the Deutsche Forschungsgemeinschaft (DFG), Grant No BI 314/23,
another (S.A.E.) received partial support from the Alexander von Humboldt foundation.
Computational time on the PL-Grid Infrastructure is gratefully
acknowledged.


\begin{thebibliography}{999}
\bibitem{1} Smith, S. B.; Finzi, L.; Bustamante, C. \textit{Science}
\textbf{1992}, \textit{258}, 1122.
\bibitem{2} Florin, E. L.; Moy, V. T.; Gaub,H. E. \textit{Science}
\textbf{1994}, \textit{264}, 415.
\bibitem{3} Tsckhvretova, L.; Trunck, J.; Sleep, J. A.; Simmons, R. M.
\textit{Nature}, \textbf{1997}, \textit{387}, 308.
\bibitem{4} Rief, M,; Gautel, M.; Oesterhelt, F.; Fernandez, J. M.; Gaub, H. E.
\textit{Science} \textbf{1997}, \textit{276}, 1109.
\bibitem{5} Essevaz-Roulet, B.; Bockelmann, U.; Heslot, F. \textit{Proc. Nath.
Acad. Sci. USA}, \textbf{1997}, \textit{94}, 11935.
\bibitem{6} Marszalek, P. E.; Oberhauser, A. F.; Pang, Y. P.; Fernandez, J. M.;
\textit{Nature}, \textbf{1998}, \textit{396}, 661.
\bibitem{7} Mehta, A. D.; Rief, M.; Spudich, J. A.; Smith, R. A.; Simmons, R. M.
\textit{Science} \textbf{1999}, \textit{283}, 1689.
\bibitem{8} Mehta, A. D.; Rock, R. S.; Rief, M.; Spudich, J. A.; Mooseker, M.
S.; Cheney, R. E. \textit{Nature} \textbf{1999}, \textit{400}, 590.
\bibitem{9} Strick, T. R.; Croquette, V.; Bensimon, D., \textit{Nature}
\textbf{2000}, \textit{404}, 901.
\bibitem{10} Kreuzer, H. J.; Grunze, M., \textit{Europhys. Lett} \textbf{2001},
\textit{55}, 640.
\bibitem{11} Bockelmann, U.; Thomen, Ph.; Essevat-Roulet, B.; Viasnoff, V.;
Heslot, F., \textit{Biophys. J.} \textbf{2002}, \textit{82}, 1537.
\bibitem{12} Holland, N. B.; Hugel, T.; Neuert, G.; Oesterhelt, D.; Moroder, L.;
Seitz, M.; Gaub, H. E., \textit{Macromolecules} \textbf{2003}, \textit{36},
2015.
\bibitem{13} Hugel, T.; Rief, M.; Seitz, M.; Gaub, H. E.; Netz, R. R.,
\textit{Phys. Rev. Lett.} \textbf{2005}, \textit{94}, 048301.
\bibitem{14} Serr, A.; Netz, R. R., \textit{Europhys. Lett.} \textbf{2006}
\textit{73}, 202. \bibitem{15} Kuhner, F.; Erdmann, M.; Gaub, H. E.,
\textit{Phys. Rev. Lett.} \textbf{2006}, \textit{97}, 21831.
\bibitem{16} Neuert, G.; Hugel, T.; Netz, R. R.; Gaub, H. E.,
\textit{Macromolecules} \textbf{2006}, \textit{39}, 789.
\bibitem{17} Strick, T. R.; Dessinges, M.-N.; Charvin, G.; Dekker, N. H.;
Allemand, J.-F.; Bennisson, D.; Croquette, V., \textit{Rep. Progr. Phys.}
\textbf{2003}, \textit{66}, 1.
\bibitem{18} Klushin, L. I.; Skvortsov, A. M., \textit{J. Phys. A: Math. Theor.}
\textbf{2011}, \textit{44}, 473001.
\bibitem{19} Subramanian, G.; Williams, D. R. M.; Pincus, P. A.,
\textit{Europhys. Lett.} \textbf{1995}, \textit{29}, 285.
\bibitem{20} Subramanian, G.; Williams, D. R. M.; Pincus, P. A.,
\textit{Macromolecules} \textbf{1996}, \textit{29}, 4045.
\bibitem{21} Williams, D. R. M.; MacKintosh, F. C., \textit{J. Phys. II
(France)} \textbf{1995}, \textbf{5}, 1407.
\bibitem{22} Gufford, M. C.; Williams, D. R. M.; Sevick, E. M.,
\textit{Langmuir} \textbf{1997}, \textit{13}, 5691.
\bibitem{23} Jimenez, J.; Rajagopalan, R., \textit{Langmuir} \textbf{1998},
\textit{24}, 2598.
\bibitem{24} Sevick, E. M.; Williams, D. R. M., \textit{Macromolecules}
\textbf{1999}, \textit{32}, 6841.
\bibitem{25} Milchev, A.; Yamakov, V.; Binder, K., \textit{Phys. Chem. Chem.
Phys.} \textbf{1999}, \textit{1}, 2083.
\bibitem{26} Milchev, A.; Yamakov, V.; Binder, K., \textit{Europhys. Lett.}
\textbf{1999}, \textit{47}, 675.
\bibitem{27} Ennis, J.; Sevick, E. M.; Williams, D. R. M., \textit{Phys. Rev.}
\textbf{1999} \textit{60}, 6906.
\bibitem{28} Skvortsov, A. M.; Klushin, L. I.; Leermakers, F. A. M.,
\textit{Europhys. Lett.} \textbf{2002}, \textit{58}, 292.
\bibitem{29} Klushin, L. I.; Skvortsov, A. M.; Leermakers, F. A. M.,
\textit{Phys. Rev. E} \textbf{2004}, \textit{69}, 061101.
\bibitem{30} Leermakers, F. A. M.; Skvortsov, A. M.; Klushin, L. I., \textit{J.
Stat. Mech.} \textbf{2004}, \textit{1}, 10001.
\bibitem{31} Skvortsov, A. M.; Klushin, L. I.; Leermakers, F. A. M., \textit{J.
Chem. Phys.} \textbf{2007}, \textit{126}, 024905.
\bibitem{32} Dimitrov, D. I.; Klushin, L. I.; Skvortsov, A. M.; Milchev, A.;
Binder, K., \textit{Eur. Phys. J.E} \textbf{2009}, \textit{29}, 9.
\bibitem{33} De Gennes, P. G. \textit{Scailing Concepts in Polymer Physics};
Cornell Univ. Press: Ithaca, N.Y. 1979. \bibitem{34} Le Guillou, J. C.;
Zinn-Justin, J., \textit{Phys. Rev.} B \textbf{1980}, \textit{21}, 3976.
\bibitem{35} Daoud, M.; De Gennes, P. G., \textit{J. Phys. (France)}
\textbf{1977}, \textit{38}, 85.
\bibitem{36} Grest, G. S.; Fetters, L. J.; Huang, J. S.; Richter, D.,
\textit{Advances Chem. Phys.} \textbf{1996}, \textit{94}, 67.
\bibitem{Egorov} Egorov, S. A.; Paturej, J.; Likos C. N.; Milchev, A.,
  \textit{Macromolecules} \textbf{2013}, \textit{46}, 3648.
\bibitem{CNL} C. N. Likos and H. M. Harreis, \textit{Cond. Matt. Phys.}, 2002,
  \textbf{5}, 173.
\bibitem{CNL1} M. Konieczny and C. N. Likos, \textit{J. Chem. Phys.}, 2006,
  \textbf{124}, 214904.
\bibitem{CNL2} M. Konieczny and C. N. Likos, \textit{Soft Matter}, 2007,
  \textbf{3}, 1130.
\bibitem{37} Halperin, A.; Alexander, S., \textit{Macromolecules} \textbf{1987},
\textit{20}, 1146.
\bibitem{38} Chen, Z.; Escobedo, F. A., \textit{Macromolecules} \textbf{2001},
\textit{34}, 8802.
\bibitem{39} Benhamou, M.; Himmi, M.; Benzoine, F.,
\textit{J. Chem. Phys.} \textbf{2003}, \textit{118}, 4759.
\bibitem{40} Paturej, J.; Milchev, A.; Egorov, S. A.; Binder, K., \textit{Soft
Matter} \textbf{2013}, {\bf DOI:} 10.1039/C3SM51275D
\bibitem{41} Sevick, E. M., \textit{Macromolecules} \textbf{2000}, \textit{33},
5743.
\bibitem{42} Fleer, G. J.; Cohen-Stuart, M. A.; Scheutjens, M. H. J.; Cosgrove,
T.; Vincent, B., \textit{Polymers at Interfaces}; Chapman and Hall: London,
1993.
\bibitem{43} Egorov, S. A., \textit{J. Chem. Phys.} \textbf{2011}, \textit{134},
194901.
\bibitem{44} Grest, G. S.; Kremer, K.: Witten, T. A.,
\textit{Macromolecules} \textbf{1987}, \textit{20}, 1376.
\bibitem{45} Grest, G. S.; Kremer, K., \textit{Phys. Rev. A} \textbf{1986},
\textit{33}, 3628.

\end{thebibliography}
\end{document}